\documentclass[preprint]{jpsj2}
\usepackage{graphicx,color,epsfig,amsmath}

\def\runauthor{Ichikawa {\sc Hirohiko}}

\title
{
Orbital Ordering in ferromagnetic Lu$_2$V$_2$O$_7$
}

\author
{ 
Hirohiko {\sc Ichikawa},
Luna {\sc Kano}, Masahiro {\sc Saitoh}, Shin {\sc Miyahara}, Nobuo {\sc Furukawa},
Jun {\sc Akimitsu}\footnote{Corresponding author e-mail: jun@phys.aoyama.ac.jp}, 
Tetsuya {\sc Yokoo}$^{1,}$,
Takeshi {\sc Matsumura}$^{2,}$, Masayasu {\sc Takeda}$^{3,}$
and 
Kazuma {\sc Hirota}$^{4,}$
}

\inst
{
Department of Physics and Mathematics, Aoyama-Gakuin University, Sagamihara 229-8558, Japan\\
$^1$Institute of Materials Structure Science, KEK, Tsukuba 305-0801, Japan\\
$^2$Department of Physics, Tohoku University, Sendai 980-8578, Japan\\
$^3$Advanced Science Research Center, JAERI, Tokai 319-1195, Japan\\
$^4$Institute of Solid State Physics, University of Tokyo, Kashiwa 277-8581, Japan
}

\recdate
{
\today
}

\abst
{
We have observed the orbital ordering in the ferromagnetic Mott-insulator
Lu$_2$V$_2$O$_7$ by the polarized neutron diffraction technique. 
The orbital ordering pattern determined from the observed magnetic 
form factors can be explained in terms of a linear combination of wave 
functions $| yz \rangle$, $| zx \rangle$ and 
$| xy \rangle$;
\begin{align*}
 | 0 \rangle &= \sqrt{\frac13}| xy \rangle
  + \sqrt{\frac13}| yz \rangle
  + \sqrt{\frac13}| zx \rangle \\
 & \propto | (x+y+z)^2-r^2  \rangle ,
\end{align*}
where each orbital is extended toward the center-of-mass of the V
tetrahedron.
We discuss the stability of the ferromagnetic Lu$_2$V$_2$O$_7$,
using a Hubbard Hamiltonian with these three orbitals.
}

\kword
{
orbital ordering, polarized neutron, Lu$_2$V$_2$O$_7$, pyrochlore
}

\begin{document}
\sloppy
\maketitle


\section{Introduction}

Vanadium pyrochlores (RE)$_2$V$_2$O$_7$, 
where RE = Lu, Yb and Tm,
exhibit ferromagnetic and semiconducting states.~\cite{rf:Bazuev}
They crystallize in a face-centered-cubic structure 
with the space group $Fd\bar{3}m$.
The crystal is characterized as a three-dimensional network
consisting of corner-sharing tetrahedra of V$^{4+}$ ions (Fig. 1).

Lu$_2$V$_2$O$_7$ ($a$ = 9.932 \AA) 
is a ferromagnetic Mott-insulator with  $T_C \simeq 73$ K. 
The electronic configuration of the V$^{4+}$ ions is $(t_{2g})^1$.
The origin of the ferromagnetism still remains unsolved.~\cite{rf:Shamoto}
In order to understand the origin,
we have to understand the electronic ground state of this system.

A systematic distortion of VO$_6$ octahedra
in a pyrochlore produces a trigonal crystal field to the vanadium ions
\cite{rf:Soderhorm}
, suggesting that the ground state is the doubly degenerate ${e_g}$ orbital.
This idea naturally leads to the existence of ``anti-ferro'' orbital ordering.
In this article we use the definition of orbital ordering in a broad sense. 
Namely, the orbital ordering means orbital polarization irrespective of its 
origin.
On the other hand, band structure calculations showed that
the non-degenerate ${a_{1g}}$ has the lowest energy state,
which is split from the ${t_{2g}}$ orbitals in a cubic crystalline field.~\cite{rf:Shamoto}

In this article, we present an orbital ordering pattern of Lu$_2$V$_2$O$_7$ 
obtained by the polarized neutron diffraction method. 



\section{Experimental}

A single crystal of Lu$_2$V$_2$O$_7$ was grown by
the floating-zone method in an Ar atmosphere.
Starting materials, Lu$_2$O$_3$ (4N) and V$_2$O$_4$ (3N),
were mixed in a given molecular ratio
and pressed into a rod. We obtained a single crystal
with 4 mm in diameter and 10 cm in length. 
The samples used in the present experiment were cut
to 4 mm in diameter and 2 mm in thickness. 

The polarized neutron diffraction measurements were performed
using the Tohoku University triple-axis spectrometer TOPAN
in the JRR-3M reactor of JAERI at Tokai.
Heusler alloy was used as a polarizer and energy of
the incident neutrons was set
at 80 meV (${\lambda}$ = 1.011 \AA) or 70 meV (${\lambda}$ = 1.081 \AA).
The beam collimation was Open-100'-60'-Open.
We also inserted a sapphire filter in front of the 2nd collimator
to eliminate neutrons with higher incident energy.

The data were taken at 3 K under an applied magnetic field of 3 T
parallel to the ${\langle 1 \ \bar{1} \ 0 \rangle}$ -axis of
the single crystal specimen.
Correction for the incomplete polarization
(neutron polarization ${P_n \sim 94.7 \ \%}$
or polarization ratio ${R \sim 17.9}$) and
estimations for the statistical uncertainty 
were properly carried out. 

We have performed similar measurements for two samples with different thickness
and found that the extinction effect is negligible.
Multiple scattering was also carefully
checked by comparing the data taken at different incident neutron energy.


\section{Results}

In order to obtain the magnetic form factor, we used a well-known relationship 
in the polarized neutron diffraction method; the observed polarization 
ratio $R$ which is the ratio of the diffracted intensities upon reversal 
of the incident neutron polarization direction, is related to 
$\gamma_0\equiv F_M/F_N$ after the instrumental corrections as
\begin{equation}
       R = \left( \frac{1 + \gamma_0 }{1 - \gamma_0 } \right) ^2, 
\end{equation}
where $F_M$ and $F_N$ are the corresponding magnetic and nuclear structure factors.
The $F_N$ values were already obtained by the x-ray diffraction.\cite{rf:Soderhorm}

Figure ~\ref{fig:1} shows the V$^{4+}$ ions in the unit cell of Lu$_2$V$_2$O$_7$.
Since we have four sites in a tetrahedron, it is possible to assume
that there are 4 types of V$^{4+}$ wave functions.
Therefore, four types of magnetic form factors $f_1$, $f_2$, $f_3$ and $f_4$ are assumed.

Using these magnetic form factors, magnetic structure factor can be expressed as 
\begin{align}
F_M &\propto \sum_{j} f_j ({\mib K}) e^{i {\mib K} \cdot {\mib r}_j } \nonumber \\
     =&     
    \left\{
      \begin{array}{clr}
      \!\!\!\! f_1 + f_2 + f_3 + f_4 & (\textrm{all } h,\, k,\, l = 4n) & (a)\\
      \left. 
         \begin{array}{c}
         f_1 - f_2 - f_3 + f_4 \\
         f_1 + f_2 - f_3 - f_4 \\
         f_1 - f_2 + f_3 - f_4
         \end{array}
      \right\}
      &
      \begin{array}{c}
        (h + k \textrm{ or }  k + l \\
        \textrm{ or } l + h = 4n) 
      \end{array}
      & (b)\\
      \!\!\!\! 0 & (\textrm{all } h,\, k,\, l = \textrm{odd}). & (c) \label{eq:form factor1}\\
      \end{array}
   \right. 
\end{align}

First, we consider the structure factors of the case $(b)$. When there is no deviation from a spherical electric charge distribution (i.e., there is no orbital ordering), the magnetic structure factor $F_M$ becomes 0 because $f_1 = f_2 = f_3 = f_4$ and no neutrons were scattered. Therefore, we can confirm the existence of the orbital ordering by observing the reflections at the Bragg points of the case $(b)$.

The results of the polarized neutron experiments are given in Table~\ref{table:1}.
In this table, ${(\mu f)_m}$ is the observed value,
and ${(\mu f)_0}$ is the value after correcting the incomplete polarization.

The final experimental results of the magnetic form factors are shown in Fig. 2 with open squares.
The bars indicate the statistical errors.
Note that the non zero values of the magnetic form factors of the (0 2 2), (4 2 2), (0 6 6), (8 2 2), (4 6 6) and (0 10 10) reflections,
corresponding to the case $(b)$ in eq.(\ref{eq:form factor1}) and shown with arrows in Fig. 2,
is a clear evidence for the existence of the orbital ordering in Lu$_2$V$_2$O$_7$.

\section{Calculation and Discussion}

We now proceed to calculate the magnetic form factors,
which depend on the aspherical spin distribution,
in order to compare the theoretically suggested orbital ordered configuration 
of the ${t_{2g}}$ electrons with the experimental results.

\subsection{Magnetic Form Factor}

The electronic configuration of V$^{4+}$ ions is $(t_{2g})^1$ and the wave 
function of the ground state can be expressed in terms of a linear combination 
of the wave functions of ${t_{2g}}$.
\begin{equation}
     | \Psi \rangle = a | yz \rangle + b | zx \rangle + 
                      c | xy \rangle ,
\end{equation}
where $| yz \rangle$, $| zx \rangle$ and 
${| xy \rangle}$ denote the wave functions of irreducible representation 
of ${t_{2g}}$.  $a$, $b$ and $c$ correspond to the relative ratio 
of the wave functions, where $a^2 + b^2 + c^2 = 1$.

The magnetic form factor $f({\mib K})$ is a quantity given by the Fourier
transformation of the magnetic moment density distribution:
\begin{equation}
       f({\mib K}) = \langle \Psi | {\mib \sigma} \cdot {\mib Q}_\bot | \Psi \rangle, \label{eq:form factor2}
\end{equation}                          
where ${\mib \sigma}$ denotes an incident neutron spin,
${\mib Q}_\bot = {\hat{\mib K}} \times {\mib Q} \times {\hat{\mib K}}$ with ${\hat{\mib K}}$ being the unit scattering vector. The scattering operator ${\mib Q}$ is given by 
\begin{equation}
      {\mib Q} = e^{i {\mib K} \cdot {\mib r}} \, {\mib s} + 
      \frac{1}{4} ({\mib l} {\mathcal{F}} + {\mathcal{F}} {\mib l}), \label{eq:Q}
\end{equation}
where ${\mib s}$ and ${\mib l}$ are the spin and the orbital
momentum operators for the electrons in V$^{4+}$ ions.
${\mathcal{F}}$ is a function defined as 
${\mathcal{F}}=\frac{2}{(i{\mib K}\cdot{\mib r})^2}
\int_0^{i{\mib K}\cdot{\mib r}} x e^x dx$ 
as given in ref. 4.
We ignore the second term in eq. (\ref{eq:Q})
because the crystalline field may quench the orbital angular momentum and
the magnetization of V$^{4+}$ ions.
Therefore $f(\mib K)$ can be expressed as
\begin{align}
   f({\mib K}) &= \langle \Psi | e^{i{\mib K}\cdot{\mib r}} s_\bot | \Psi \rangle \nonumber \\
         &= \langle j_{0} \rangle  - \frac{5}{14}\bigl[(a^2 +b^2 -2c^2)(3\cos^2\theta - 1) \nonumber \\
         & {} \qquad\, + 12c\sin\theta\cos\theta(a\cos\phi + b\sin\phi) \nonumber \\
         & {} \qquad\, - 3\sin\theta\{(a^2 - b^2)\cos2\phi - 
               2ab\sin2\phi\}\bigr] \langle j_{2} \rangle \nonumber \\
         & {} - \frac{3}{56}\bigl[(4a^2 + 4b^2 - c^2)(35\cos^4\theta - 
               30\cos^2\theta + 3) \nonumber \\
         & {} \quad\, + 20c\sin\theta(7\cos^3\theta - 3\cos\theta) 
               (a\cos\phi + b\sin\phi) \nonumber \\
         & {} \quad\, + 20\sin^2\theta(7\cos^2\theta - 1)  \nonumber \\
         & {} \qquad\, \times  \{(a^2 - b^2) \cos2\phi - 2ab\sin2\phi\} \nonumber \\
         & {} \quad\, + 140c\sin^3\theta\cos\theta(a\cos3\phi - 
               b\sin3\phi) \nonumber \\
         & {} \quad\,                + 35c^2\sin^4\theta\cos4\phi\bigr]
               \langle j_{4} \rangle ,
\end{align}
where $\theta$ and $\phi$ denote the spherical coordinates of the scattering 
vector ${\mib K}$ relative to the quantization axes. 
$\langle j_{n} \rangle$ ($n$ = 0, 2 and 4) are the Freeman-Watson radial functions \cite{rf:Watson},
which are calculated using the program cited in ref. 6.

\subsection{Orbital Ordered Structure in ${\rm Lu_2V_2O_7}$}
To reproduce the magnetic form factor observed 
in ${\rm Lu_2V_2O_7}$, we have made model calculations 
on possible two types of models for orbital ordering structures.
One is a model with three-fold degenerate orbitals
in a tetragonal crystal field (model A) and the other is
a model in a trigonal crystal field due to an oxygen lattice
distortion (model B).

As shown in Fig.~1, the unit cell of ${\rm Lu_2V_2O_7}$
is constructed from a tetrahedron of 
four corner-shared ${\rm VO_6}$ octahedra, 
and each ${\rm V^{4+}}$ ($3 d^1$) ion has
a three-fold degenerate $t_{2g}$ orbital
in the tetragonal crystal field. 
The hopping amplitude of the electon depends on the direction 
of the hoppings and the orbital states
due to the symmetries and anisotropies of the $3d$ orbitals.
We define the local quantization axis 
on each ${\rm V^{4+}}$ as shown in Fig.~3 (a) and 
express the $t_{2g}$ orbital states on the site $i$
as $| xy \rangle_i$, $| yz \rangle_i$, and $| zx \rangle_i$.
Geometrically, the $t_{2g}$ orbital on each site
is approximately parallel to the surface of 
the V$^{4+}$ tetrahedron.
Since the V$^{4+}$ ions structure is constructed with the 
corner-shared octahedra, the largest contribution for
the $t_{2g}$ electron hopping is in general considered 
to be from $t_{2g}$-$p_{\pi}$-$t_{2g}$ couplings.~\cite{kanamori59}  
Let us focus on the electron hoppings between
the site $1$ and $2$ (Fig.~3 (a)).
For example, the $|yz \rangle_1$
and $|zx \rangle_2$ orbitals are roughly 
on the surface $123$ in Fig.~3 (b).
If the extents of the rotation and tilting were not very large,
electron transfers between the orbitals parallel to
the surface would be much larger than the other components.
In fact, if the V-O-V angle were $\theta = 180^{\circ}$,
there would be a hopping process $t_1$ from a $t_{2g}$-$p_{\pi}$-$t_{2g}$
coupling between $|yz \rangle_1$ and $|zx \rangle_2$.
In the same way, the electron hopping from $|zx \rangle_1$ to 
$|yz \rangle_2$ is also favorable,
because these orbitals are approximately parallel to 
the triangle $124$ (see Fig.~3 (c)).
On the other hand, since the direction from site 1 to 2
is nearly perpendicular to both the $|xy \rangle_1$ and $|xy \rangle_2$, 
the hopping amplitude from these orbitals
should be small (see Fig.~3 (d)).
In this way, there are two dominant hopping processes on each bond.
Strictly speaking, the bond angle $\theta$
in ${\rm Lu_2V_2O_7}$ is smaller than $180^{\circ}$ and
there are two kinds of hopping processes between $t_{2g}$ orbitals
$t_1$ and $t_2$ as shown in Fig.~\ref{fig:process} (a).
Nevertheless, since the angular dependence of the hoppings are
$t_1 \propto \cos\theta - 1$ and $t_2 \propto \cos\theta + 1$,
we can neglect the hopping process $t_2$ as a 
first order approximation. 



Let us consider the model A (see Fig.~\ref{fig:process} (a)) 
based on the Hubbard Hamiltonian 
with three-fold degenerate orbitals in the limit $t_1/U \ll 1$,
where we only take into account the most dominant hopping process $t_1$.
Since the electron in the $| yz \rangle_1$ orbital
can move only within the hatched surface 
in Fig.~\ref{fig:hopping} (a), 
i.e., the triangle $123$ of the tetrahedron, 
only one electron can exist in the three orbital states 
$| yz \rangle_1$, $| zx \rangle_2$, and $| xy \rangle_3$ 
on the surface-plane to gain the kinetic exchange energy.
Thus due to the Pauli principle it is expected that 
the energy gain is maximum when the $3d$ electrons 
on ${\rm V^{4+}}$ ions occupy the $t_{2g}$ orbitals 
which lie on different surface-planes so that the virtual 
hoppings of electrons $t_1$ are not prohibited. 
An example of such an orbital ordering is realized
when all occupied orbital states on the ${\rm V^{4+}}$ ions 
can simultaneously be expressed as one orbital, e.g., 
$| yz \rangle$ orbital. Namely
a ``ferro''-orbital ordered state should be favored.
Note that the orbitals on each site face
the different directions even in the ``ferro''-orbital ordered state
since we define the local quantization axis on each site differently.
  
Next, we consider the model B (see Fig.~\ref{fig:process} (b)).
In the trigonal crystal field,
the three-fold degeneracy of $t_{2g}$ orbitals is lifted.
The lowest energy state which has an $a_{1g}$ symmetry is 
written as,
\begin{equation}
  | 0 \rangle = 
  \frac{1}{\sqrt{3}} (| xy \rangle + | yz \rangle
  + | zx \rangle).
  \label{eq:trigonal_base_g}
\end{equation}
The orbital ordering structure is shown in Fig.~\ref{fig:o-o-a1g}, 
where each orbital is extended toward the center-of mass of the
tetrahedron.


Now we compare the model A and B.
It is expected that there is a competition 
between kinetic energy gain through the orbital exchange 
and splitting energy gain due to the trigonal crystal field, 
which determine the orbital structure in ${\rm Lu_2V_2O_7}$. 
Assuming the orbital states in the models,
we calculate the magnetic form factors and
compare the results with that observed in the experiment.
As shown in Fig.~2, the form factors in the model B
agree much better with the experimental results.
The method of least squares indicates that 
``ferro''-orbital ordered state with 
\begin{equation}
  | \phi \rangle_i = a | xy \rangle_i + b | yz \rangle_i 
  + c | zx \rangle_i,	
\end{equation}
where $a = 0.67 \pm 0.25$, $b = 0.40 \pm 0.23$, 
$c = 0.62 \pm 0.14$,
is favored. 
The orbital state $| \phi \rangle_i$ agrees 
with that observed in model B ($a=b=c=0.58$)
within the computed error bars. 
Thus we conclude that the orbital ordered structure
shown in Fig.~\ref{fig:o-o-a1g} is realized in ${\rm Lu_2V_2O_7}$.
Our orbital structure model is also consistent with that proposed 
by Shamoto {\it et al.}~\cite{rf:Shamoto}
In this way, it is expected that 
the effects of the trigonal field is bigger than
that of the exchange energy of the electrons  in ${\rm Lu_2V_2O_7}$. 
In fact, the temperature $T_{oo}$ at which orbital ordering occurs
is much higher than the Curie temperature $T_c = 73$K.  
  
Finally we check the stability of the ferromagnetic state,
which has been observed in the experiment, on the model B.  
Following the procedure in Ref.~\citen{kugel73},
we start from the Hubbard Hamiltonian with $3d$ orbitals:
\begin{align}
  H & = \sum_{i,j} \sum_{m, m', \sigma} t^{m, m'}_{i, j} 
  c^{\dagger}_{i, m, \sigma} c_{j, m', \sigma}  \nonumber \\
  & + U \sum_{i, m} n_{i m \uparrow} n_{i m \downarrow}
  + U^{\prime} \sum_{i, \sigma, m \neq m' } n_{i m \sigma} 
  n_{i m' -\sigma} \nonumber \\
  & + ( U^{\prime} - J_H ) \sum_{i, \sigma,  m \neq m'}  n_{i m \sigma}  n_{i m' \sigma} \nonumber \\
  & -J_H \sum_{i,  m \neq m'} (c^{\dagger}_{i m \uparrow} 
  c_{i m \downarrow} c^{\dagger}_{i m' \downarrow} c_{i m' \uparrow} \nonumber \\
  & \qquad\qquad\qquad + c^{\dagger}_{i m \uparrow} c^{\dagger}_{i m \downarrow}
  c_{i m' \uparrow}  c_{i m' \downarrow} + {\rm H.c.} ) \nonumber \\
  & + \sum_{i, m, \sigma} \Delta_m n_{i m \sigma}.
  \label{eq:three-band-Hubbard}
\end{align}
Here $U$ is an intra-orbital Coulomb interaction,
$U^{\prime}$ an inter-orbital Coulomb interaction, 
$J_H$ Hund's rule coupling, and $\Delta_m$ 
splitting energy due to the trigonal crystal field.
Indices $m, m'$ indicate orbital states,
$i, j$ sites, and $\sigma$ spin states. 

Under the trigonal crystal field,
the three-fold degeneracy of the $t_{2g}$ orbital
is lifted: the lowest energy state $| 0 \rangle$ 
which is defined in eq.(\ref{eq:trigonal_base_g}) 
and the excited states with the $e_{g}$ symmetry.
We define that the splitting energy for the lowest energy state
$|0 \rangle$ is zero, and that for the excited state
$\Delta_0$ in the Hamiltonian (\ref{eq:three-band-Hubbard}). 
Given the transfers $t_1^{\prime}$ and $t_2^{\prime}$
as shown in Fig~\ref{fig:process} (b),
we can obtain an effective Hamiltonian 
in the second order perturbation for $t_1^{\prime}/U$
and $t_2^{\prime}/U$,
\begin{align}
  H_{eff} & =  
  - \left\{ \frac{16 t_1^{\prime 2}}{9 U} 
        + \frac{8 t_2^{\prime 2}}{9 U^{\prime}}
  \left( 1 - \frac{J_H}{U^{\prime}} \right)  
  \right\} \left( \frac{1}{4} - {\bf s}_1 \cdot {\bf s}_2 \right)        
  \nonumber \\
  &  - \frac{8 t_2^{\prime 2}}{9 U'} 
  \left( 1 + \frac{J_H}{U^{\prime}} \right)  
  \left(\frac{3}{4} + {\bf s}_1 \cdot {\bf s}_2 \right),
  \label{eq:H_eff}
\end{align}
where we assume the condition that $U, U^{\prime} \gg J_H \gg \Delta_0 = 0$.
When the spins of the electrons on site 1 and site 2 are antiparallel,
the energy gain through the virtual hopping process
to the intermediate states $|0 \rangle$ and $| \pm \rangle$
are reduced by the first term of 
eq.(\ref{eq:H_eff}). On the other hand, 
when the spins are parallel, the transfer $t_1$ is forbidden
by Pauli's principle and the energy gain through 
the intermediate states $| \pm \rangle$ are possible by
the transfers $t_2^{\prime}$ as given in the second
term of eq.(\ref{eq:H_eff}).
We have checked the stability of the ferromagnetic state,
assuming that $U = 6.0$ eV.~\cite{rf:Shamoto}   
The result indicates that the ferromagnetic state is
stable for $J_H \gtrsim 1.8$ eV, where we use
the relation $|t_2^{\prime}| = |t_1^{\prime}|/2$ (see Appendix).
Hund's rule couplings estimated in ${\rm Lu_2V_2O_7}$
$J_H = 1.5$ eV ($U^{\prime} = 3.0$)\cite{rf:Shamoto}
and $J_H = 0.7$ eV ($U^{\prime} = 4.6$ eV)\cite{mizokawa03} 
are slightly smaller than the critical value $1.8$ eV.
This may contradict the experimental results. 

So far we take into account the hopping processes
between $t_{2g}$ orbitals. However, 
including the effects of the hopping process 
from $t_{2g}$ orbitals to $e_g$ orbitals
($|x^2-y^2 \rangle$ and $|3 z^2 - r^2 \rangle$)
might be important to stabilize the ferromagnetism
in $t_{2g}$ orbital system (see fig.4(b)), 
as predicted by Mochizuki and Imada
in the Perovskite-type Ti oxides.~\cite{Mochizuki00}
In fact, even in ${\rm Lu_2V_2O_7}$, 
the ratio of the hopping processes
for the V-O-V angle $\theta = 130^{\circ}$ are
$|t_3^\prime|/|t_1^\prime| \sim 3$ (see Appendix).
Therefore the effects of the virtual hopping process $t_3^\prime$ 
can not be neglected even if 
there is a level splitting $\Delta_1$ due to the crystal field.
Including the effects of $t_3^\prime$,
the second order effective Hamiltonian
is written as 
\begin{align}
  H_{eff} & = \sum_{i,j} J(J_H, \theta, \Delta_0, \Delta_1)
  {\bf s}_i \cdot {\bf s}_j
  + {\rm const}.  
  \label{eq:H_eff_eg}
\end{align}
$J(J_H, \theta, \Delta_0, \Delta_1)$ has been calculated 
as a function of $J_H$ and $\Delta_1$ with fixed values
$\theta = 130^{\circ}$ and $\Delta_0 = 0.4$ eV.
The parameter ranges where the ferromagnetic state 
is stable are shown in Fig.~\ref{fig:6}.
In this way, the energy gain through the 
virtual hopping process $t^{\prime}_3$
can stabilize the ferromagnetism rather than 
the antiferromagnetism.
It is natural to think that in  ${\rm Lu_2V_2O_7}$
the virtual hopping process $t^{\prime}_3$
can play an important role to stabilize the ferromagnetism,
since the ferromagnetic ground state 
has been observed experimentally in this material.
In fact, assuming the level splitting 
energy $\Delta_1 = 2.0$ eV\cite{mizokawa03},
both estimated values of Hund's rule coupling
$J_H = 1.5$ eV\cite{rf:Shamoto} and $J_H = 0.7$ eV\cite{mizokawa03}
indicate the presence of the ferromagnetic ground state in ${\rm Lu_2V_2O_7}$, 
which is consistent with the experimental results. 
Once the V-O-V bond angle is away from $180^{\circ}$, the transfer
amplitude between $t_{2g}$ to $e_{g}$ orbitals can not be negligible.
As a result, the ferromagnetic state can be favorable due to the 
energy gain by the second order perturbation processes 
through the $t_2$ and $t_3$ hopping transfers.


\section*{Acknowledgments}
We would like to acknowledge Y. {\sc Tokura} and Y. {\sc Taguchi} 
for informing us of this problem and 
also S. {\sc Shamoto} and T. {\sc Mizokawa}
for many helpful discussions. This work was supported by 
a Grant-in-Aid for Aoyama Gakuin University 21st COE 
Program from the Ministry of Education, Culture, Sports,
Science and Technology of Japan.

\section*{Appendix}  
Under the tetragonal crystal field,
a five-fold degeneracy of $3d$ orbitals
on ${\rm V^{4+}}$ ($3 d^1$) ion is lifted 
to three-fold lower levels $t_{2g}$ state and
two-fold higher levels $e_g$ state 
(see Fig.~\ref{fig:process} (a)).
Because of symmetries and anisotropies of the $3d$ orbitals,
the hopping amplitude depends on the direction 
of the hoppings and the orbital states.
Let us consider the hopping process 
from $|yz \rangle_1$ to the orbitals on the site 2.
Since the V-O-V bond angle 
in ${\rm Lu_2V_2O_7}$ is about $\theta = 130^{\circ}$, 
the $|yz \rangle_1$ orbital hybridizes with 
not only $|zx \rangle_2$ but also
$|yz \rangle_2$ and $|3 z^2 - r^2 \rangle_2$. Thus 
hopping amplitudes $t_2$ and $t_3$ defined in 
Fig.~\ref{fig:process} (a) are non-zero.
The angle dependences of the hoppings are written as
\begin{align} 
  t_1 & = -\frac{1}{2 \Delta} 
  V_{pd\pi} V_{pd\pi} (\cos\theta - 1), \\
  t_2 & = -\frac{1}{2 \Delta} 
  V_{pd\pi} V_{pd\pi} (\cos\theta + 1), \\
  t_3 & = -\frac{1}{\sqrt{2} \Delta} 
  V_{pd\pi} V_{pd\sigma} \sin\theta,
\end{align}
where $\Delta$ is the charge-transfer energy
which describes the energy difference between
occupied O $2p$ and unoccupied V $3d$ levels,
and $V_{pd\sigma}$ and $V_{pd\pi}$ are 
interatomic matrix elements for $\sigma$ and $\pi$
bonds, respectively.

Under the trigonal crystal field,
the degeneracy of $t_{2g}$ states is lifted.
The lowest state
$|0 \rangle$ is defined in eq.(\ref{eq:trigonal_base_g}) and
two states with the $e_{g}$ symmetry $|+ \rangle$ and
$|- \rangle$ are
\begin{align}
  | + \rangle & = - \frac{1}{\sqrt{3}} 
  (| xy \rangle + \omega | yz \rangle
  + \omega^2 | zx \rangle), \\
  | - \rangle & = \frac{1}{\sqrt{3}} 
  | xy \rangle + \omega^2 | yz \rangle
  + \omega | zx \rangle),
\end{align}
where $\omega \equiv \exp(2 \pi i / 3)$.
Note that the higher $e_g$ orbitals $| x^2 - y^2 \rangle$ and
$| 3 z^2 - r^2 \rangle$ remain unlifted.
Using the hopping processes $t_1$, $t_2$, and $t_3$,
possible hopping processes shown in Fig.~\ref{fig:process} (b)
have been calculated and the results are
\begin{align}
  t_1^{\prime} & = -\frac{2}{3 \Delta} 
  V_{pd\pi} V_{pd\pi} \cos\theta, \\
  t_2^{\prime} & = \frac{1}{3 \Delta} 
  V_{pd\pi} V_{pd\pi} \cos\theta, \\
  t_3^{\prime} & = -\frac{\sqrt{2}}{\sqrt{3} \Delta} 
  V_{pd\pi} V_{pd\sigma} \sin\theta.
\end{align}
The ratio of interatomic matrix element $V_{pd\sigma}$ 
and $V_{pd\pi}$ is fixed at 
$V_{pd\sigma}/V_{pd\pi} \cong -2.15$,
which is the value obtained in the Perovskite compounds.~\cite{Harrison}
Using the V-O-V angle $\theta = 130^{\circ}$,
the ratio of the hopping processes are
$|t_1^\prime| : |t_2^\prime| : |t_3^\prime| \sim 2 : 1 : 6$.


\pagebreak
\noindent
Figure captions
\\
\

\noindent
Figure.\ref{fig:1}
{V$^{4+}$ ions in the unit cell of (RE)$_2$V$_2$O$_7$ (RE = Lu, Yb and Tm).}
\\
\

\noindent
Figure.\ref{fig:2}
{Comparison between the observed and the calculated magnetic 
        form factors. The open squares with bars correspond to the experimental values. 
        The open circles are the calculated values of the orbital ordering 
        model A, and the filled circles are the ones of the model B.
        The computed error bars of the least squares calculation are within plots.
}
\\
\

\noindent
Figure.\ref{fig:hopping}
{(a) The definition of the directions of the $x, y, z$ 
    axes at each $V$ site on a tetrahedron. Circles represent
    O ions.
    (b) Orbital states on the site 1 and 2,
    which are approximately parallel to the hatched plane $123$.
    The $t_{2g}$ electron hoppings through
    $t_{2g}$-$p_{\pi}$-$t_{2g}$ couplings
    are shown by arrows
    ($p$ orbitals are omitted for simplicity).
    (c) Orbital states on the site 1 and 2,
    which are approximately parallel to the plane $124$.
    The $t_{2g}$ electron hoppings are shown by arrows.
    (d) An orbital state on the site 1 (2) which is 
    approximately parallel to the plane $134$ ($234$).
    The $t_{2g}$ electron in these orbitals 
    can not hop from site 1 to site 2.
    } 
\\
\

\noindent
Figure.\ref{fig:process}
{Characteristic hopping process.
    (a) in a tetragonal crystal field.
    (b) in a trigonal crystal field. }
\\
\

\noindent
Figure.\ref{fig:o-o-a1g}
{Schematic view of the orbital ordering
    in Lu$_2$V$_2$O$_7$ (model B).}
\\
\

\noindent
Figure.\ref{fig:6}
{The parameter region where ferromagnetic 
    (antiferromagnetic) interaction couplings are realized.
    Estimated values by Shamoto {\it et al.} and Mizokawa
    are shown by circles.}

\pagebreak
\begin{fullfigure}
\begin{center}
\epsfig{file=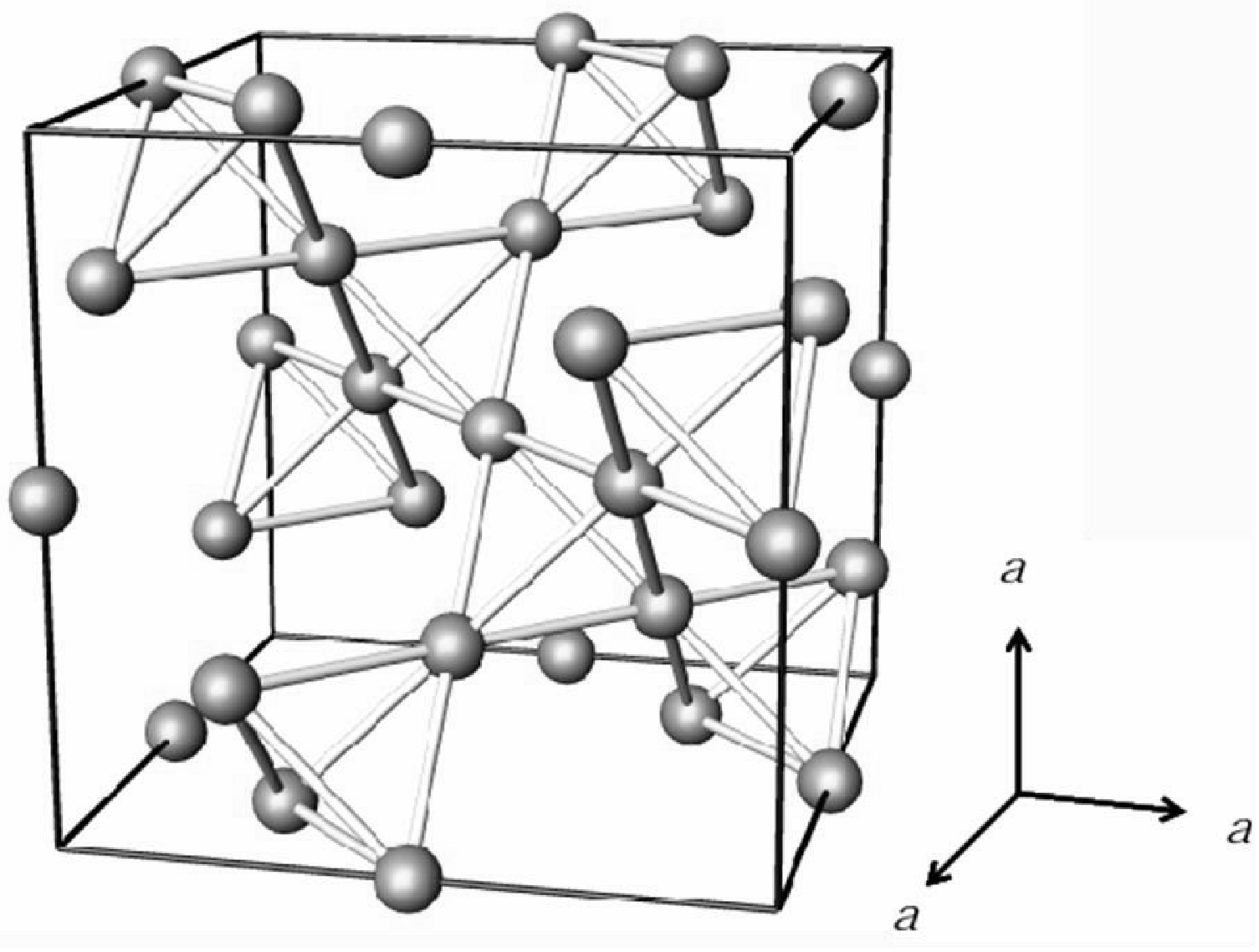, width=0.8 \textwidth}
\end{center}
\caption{}
\label{fig:1}
\end{fullfigure}

\begin{flushright}
H. Ichikawa \it{et. al.}
\end{flushright}

\pagebreak
\begin{fullfigure}
\begin{center}
\epsfig{file=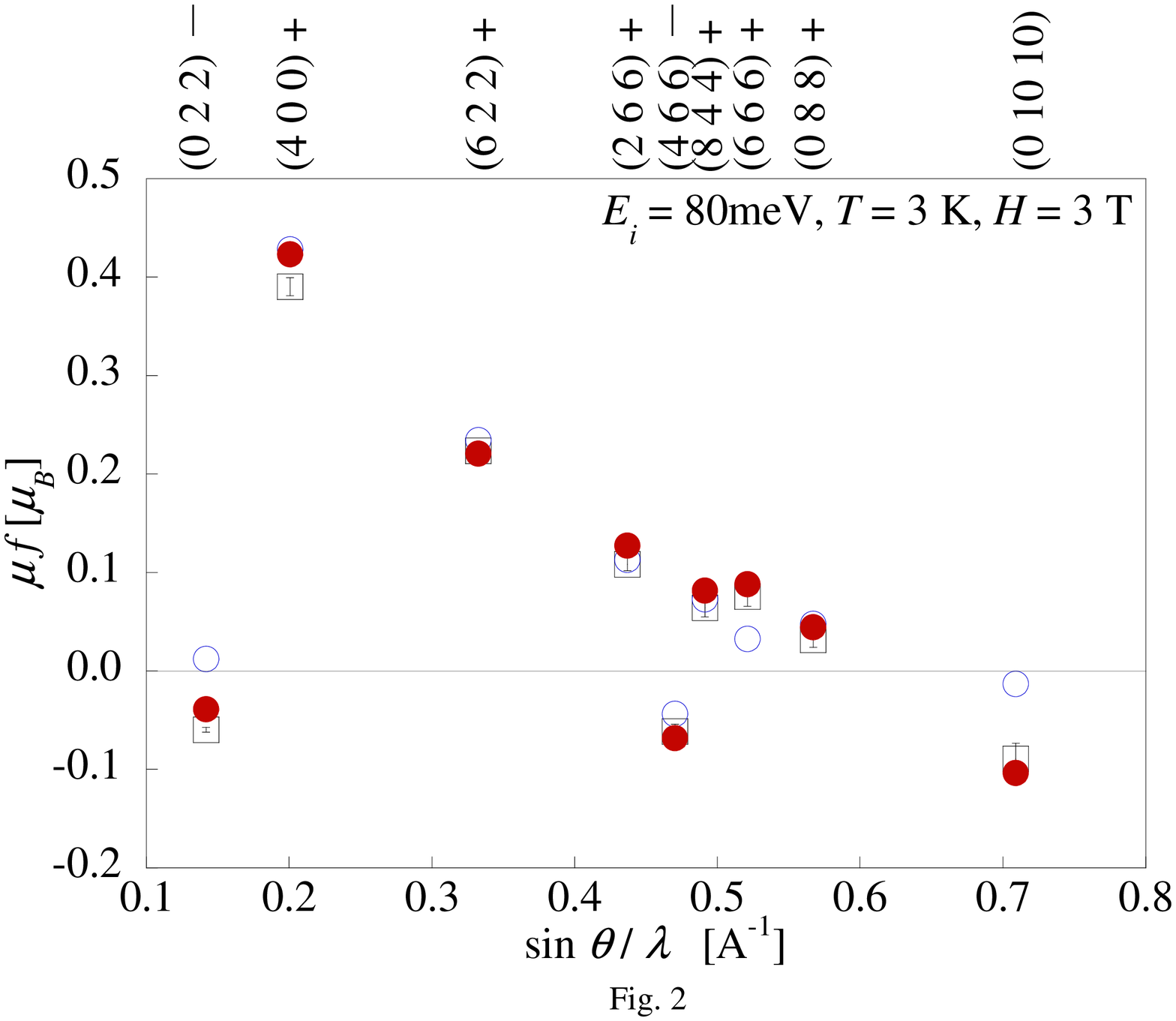, width=0.8 \textwidth}
\end{center}
\caption{}
\label{fig:2}
\end{fullfigure}

\begin{flushright}
H. Ichikawa \it{et. al.}
\end{flushright}

\pagebreak
\begin{fullfigure}
  \begin{center}
    \epsfig{file=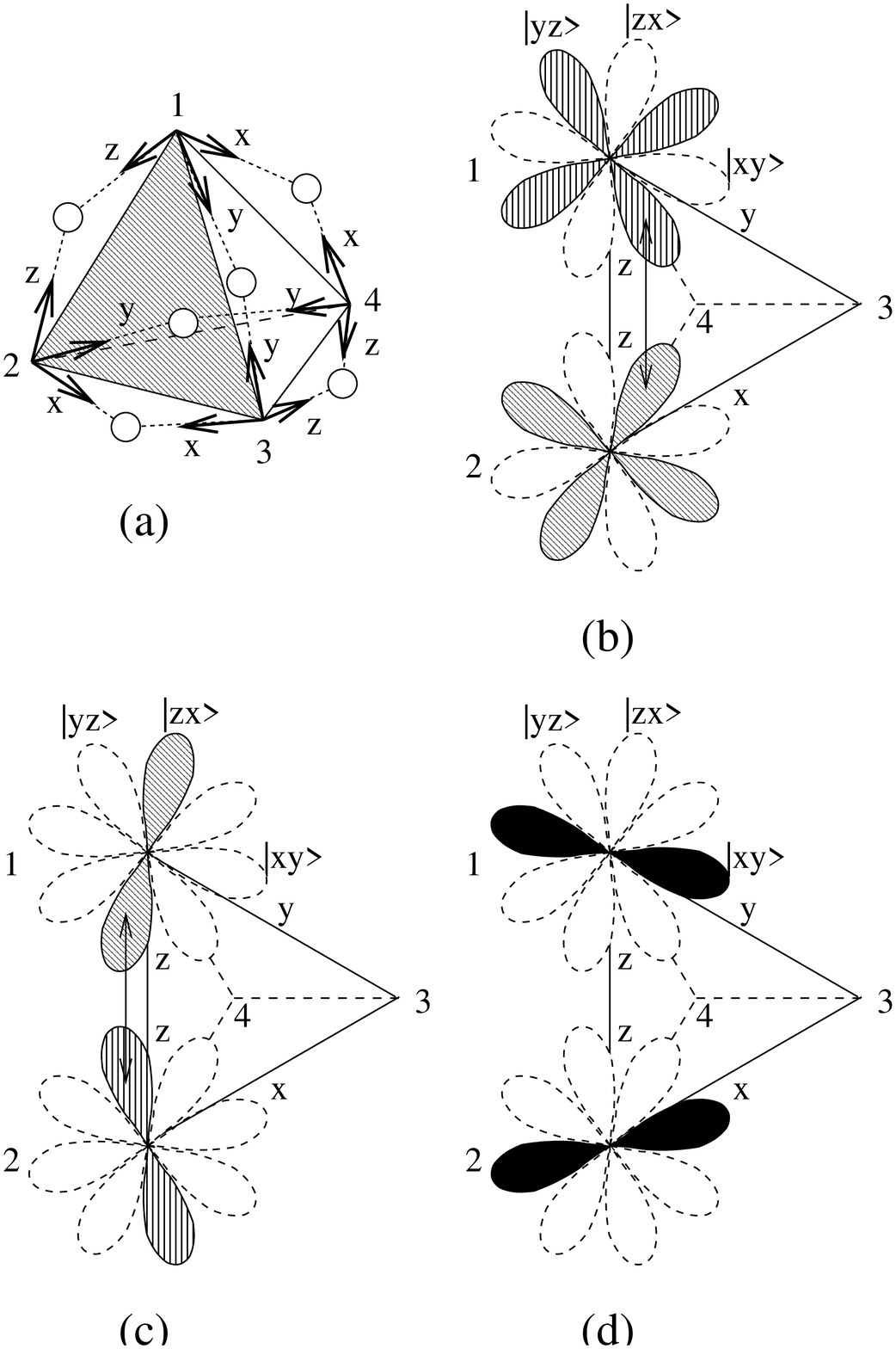, width=0.65 \textwidth}
  \end{center}
\caption{}
  \label{fig:hopping}
\end{fullfigure}

\begin{flushright}
H. Ichikawa \it{et. al.}
\end{flushright}

\pagebreak
\begin{fullfigure}
  \begin{center}
    \epsfig{file=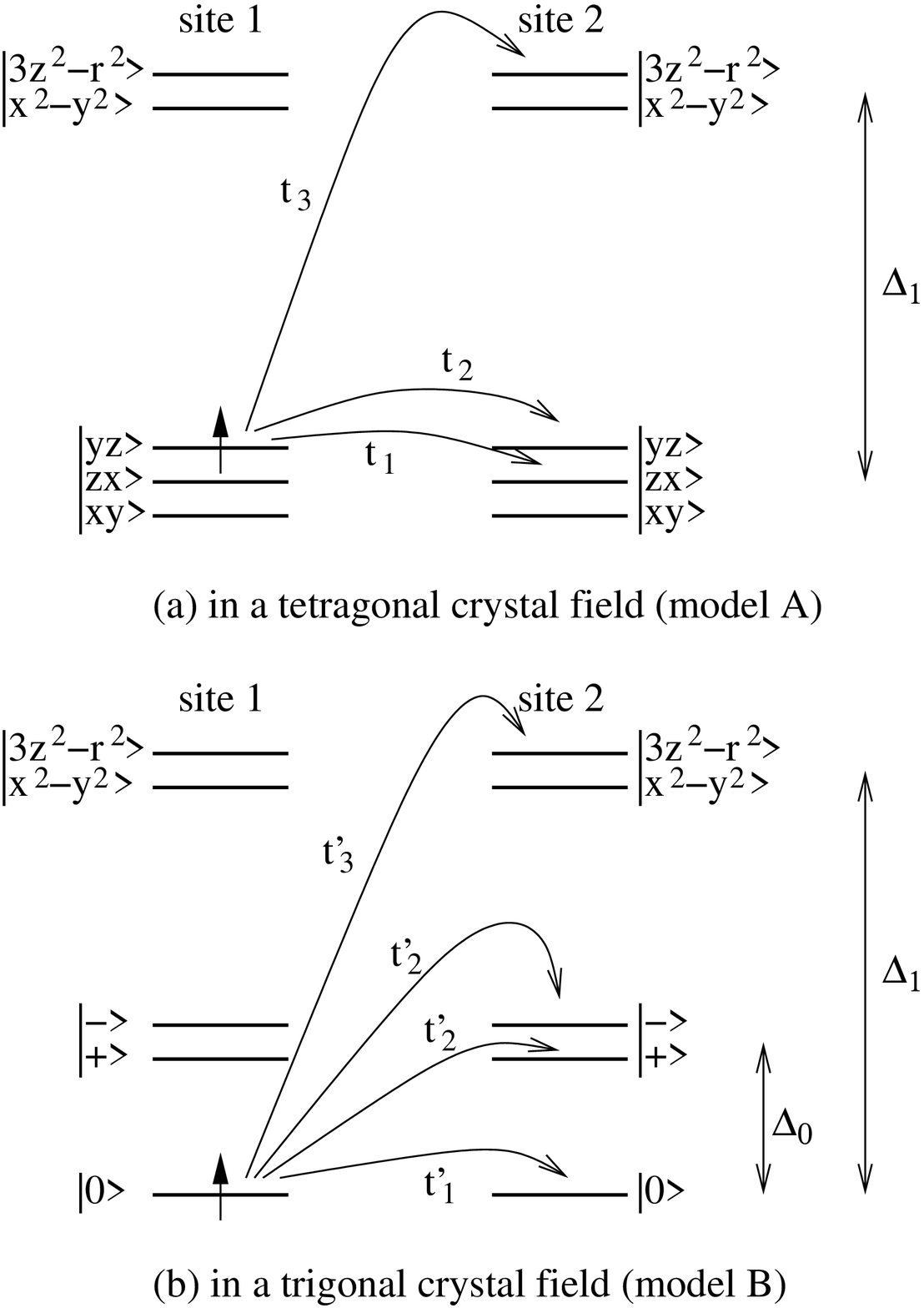, width=0.65 \textwidth}
  \end{center}
\caption{}
  \label{fig:process}
\end{fullfigure}

\begin{flushright}
H. Ichikawa \it{et. al.}
\end{flushright}

\pagebreak
\begin{fullfigure}
  \begin{center}
    \epsfig{file=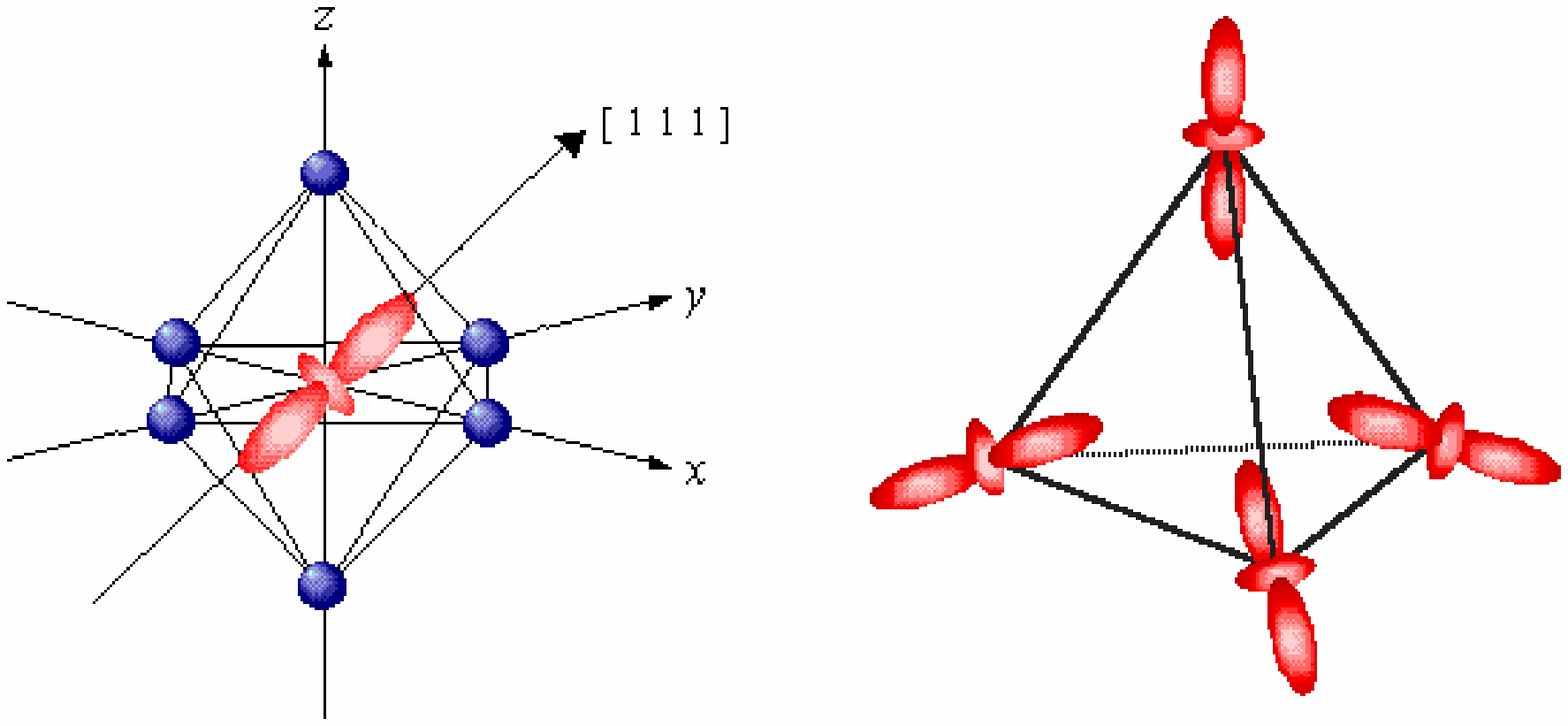, width=0.8 \textwidth}
  \end{center}
\caption{}
  \label{fig:o-o-a1g}
\end{fullfigure}

\begin{flushright}
H. Ichikawa \it{et. al.}
\end{flushright}

\pagebreak
\begin{fullfigure}
  \begin{center}
    \epsfig{file=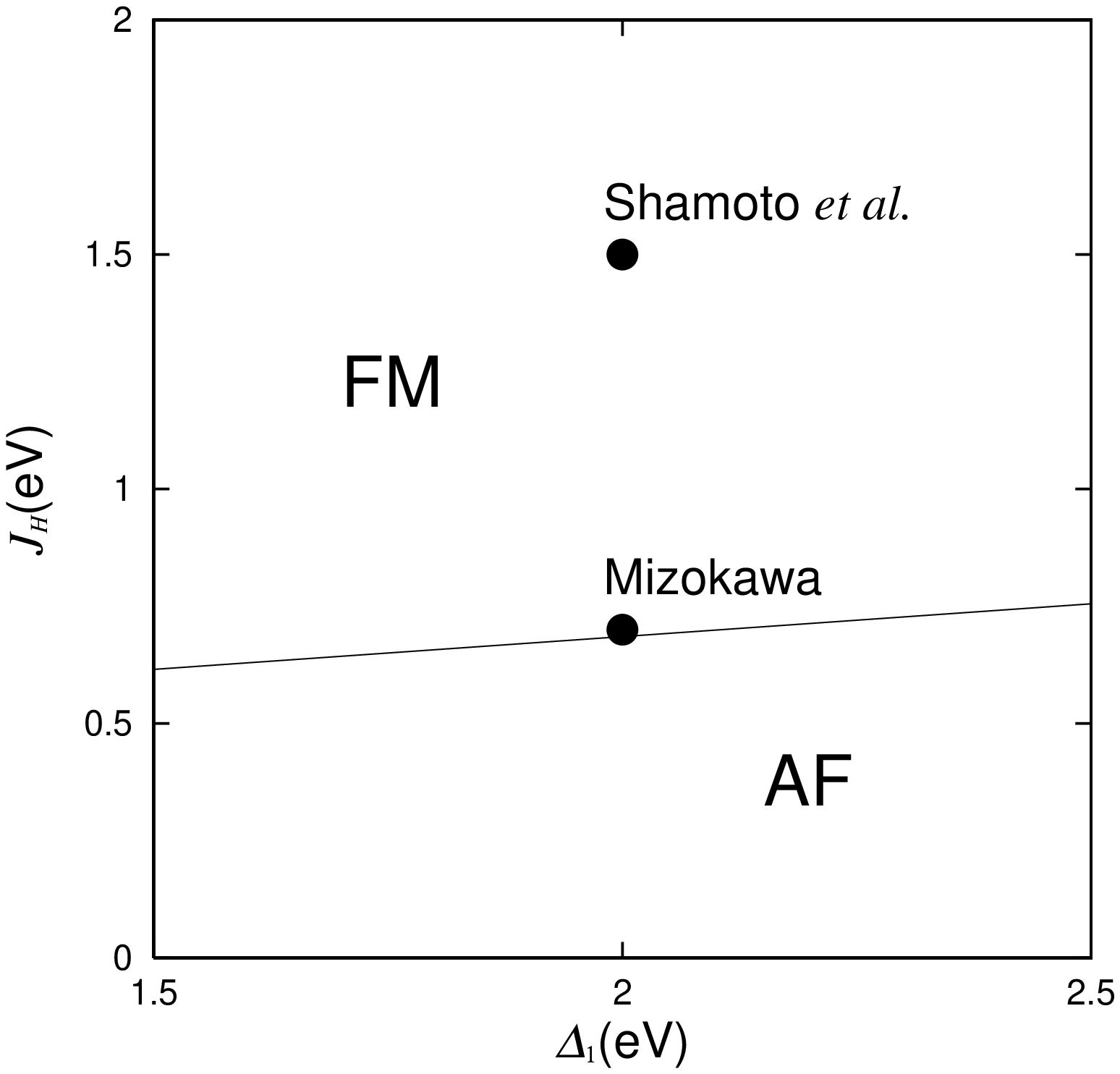, width=0.8 \textwidth}
  \end{center}
\caption{}
  \label{fig:6}
\end{fullfigure}

\begin{flushright}
H. Ichikawa \it{et. al.}
\end{flushright}

\pagebreak
\begin{fulltable}
\caption{Comparison between the experimental and calculated data.

${(\mu f)_m}$ are the observed values, 
${(\mu f)_0}$ are the observed values after correction, 
${(\mu f)_A}$, ${(\mu f)_B}$ are the calculated values,
corresponding to the model A and B with $\mu = 0.63\ \mu_B$.
Parenthesis are the statistical errors.
}

\label{table:1}
\begin{fulltabular}{@{\hspace{\tabcolsep}\extracolsep{\fill}}ccrrrr} 
\hline
$(h\ k\
 l)$ & ${\sin\theta / \lambda}$ (\AA$^{-1}$) & ${(\mu f)_{m} \quad }$ 
 & ${(\mu f)_{0} \quad\, }$ & ${(\mu f)_{A}}$ & ${(\mu f)_{B}}$ \\ \hline

(0 2 2)   & 0.1424  & -- 0.055(2) \, & -- 0.062(2) \, & -- 0.001  & -- 0.038 \\
(4 0 0)   & 0.2014  &    0.358(2) \, &    0.401(3) \, &    0.399  &    0.409 \\
(4 2 2)   & 0.2466  & -- 0.022(12)   & -- 0.025(13)   &    0.006  & -- 0.011 \\
(6 2 2)   & 0.3339  &    0.212(5) \, &    0.238(5) \, &    0.207  &    0.210 \\
(4 4 4)   & 0.3488  &    0.204(12)   &    0.229(13)   &    0.177  &    0.198 \\
(0 6 6)   & 0.4272  & -- 0.127(9) \, & -- 0.142(10)   & -- 0.015  & -- 0.136 \\
(8 2 2)   & 0.4272  &    0.013(13)   &    0.014(14)   &    0.009  &    0.011 \\
(2 6 6)   & 0.4389  &    0.107(5) \, &    0.120(5) \, &    0.093  &    0.110 \\
(4 6 6)   & 0.4723  & -- 0.058(6) \, & -- 0.065(6) \, &    0.004  & -- 0.063 \\
(8 4 4)   & 0.4933  &    0.063(8) \, &    0.070(8) \, &    0.058  &    0.074 \\
(6 6 6)   & 0.5232  &    0.077(8) \, &    0.086(8) \, &    0.032  &    0.067 \\
(0 8 8)   & 0.5696  &    0.029(6) \, &    0.032(7) \, &    0.023  &    0.038 \\
(4 8 8)   & 0.6041  &    0.034(11)   &    0.038(12)   &    0.006  &    0.036 \\
(0 10 10) & 0.7119  & -- 0.096(14)   & -- 0.107(15)   & -- 0.022  & -- 0.117 \\
\hline
\end{fulltabular}
\end{fulltable}

\begin{flushright}
H. Ichikawa \it{et. al.}
\end{flushright}

\end{document}